\documentclass{elsart}
\usepackage{makeidx}
\usepackage{graphicx}
\usepackage{multicol}
\usepackage{times}
\usepackage{cite}
\usepackage{amsmath}
\usepackage{amssymb}

\makeindex

\begin{document}

\title{Mixing, Ergodicity and slow relaxation phenomena}
\author{I.V.L. Costa,}
\author{M.H. Vainstein,} 
\author{L.C. Lapas,}
\author{A.A. Batista,} 
\author{F.A. Oliveira}
\maketitle

\address{Instituto de F\'{\i}sica and International Center for Condensed
Matter Physics, Universidade de Bras\'{\i}lia, CP 04513, 70919-970,
Bras\'{\i}lia-DF, Brazil \texttt{fao@fis.unb.br}}

\small
\textbf{ABSTRACT} Investigations on diffusion in systems with
 memory~\cite{Costa03} have established a hierarchical connection
between mixing, ergodicity, and the fluctuation-dissipation
theorem (FDT). 
This hierarchy means that ergodicity is a necessary condition for the
validity of the FDT, and mixing is a necessary condition for ergodicity.
In this work, we compare those results with recent
investigations  using the  Lee recurrence
relations method~\cite{Lee82,Lee01,Lee06}. 
Lee shows that ergodicity is violated in the dynamics of the electron gas~\cite{Lee06}. 
This reinforces both works and implies that the results of~\cite{Costa03} are more general than the framework in which they were obtained. Some
applications to slow relaxation phenomena are discussed. 

\emph{Introduction --} More than a hundred years after its formulation by
Boltzmann, the ergodic hypothesis (EH) still drives the
attention of the mathematics and physics community. Many situations have been
found in which the EH is broken~\cite{Lee01,Costa03,Lee06,Mukamel05}. On the
other hand, the mixing condition (MC), despite being omnipresent in relaxation
processes, has seldomly been the object of investigation~\cite
{Costa03,Vainstein06}. Recent studies on anomalous diffusion~\cite
{Morgado02,Costa03} have established a strong hierarchy, which in growing
generality is: the FDT, EH, and the MC. Since the FDT is a less general
``theorem'', we shall focus only on the EH and the MC. 

Kubo~\cite{Kubo57} realized that doing a time average on correlation functions
would be more realistic than performing those averages on the variables
themselves. He then established an ergodic condition in the framework of
linear-response theory. 
That is, \textit{if the zero-frequency limit of a dynamical
susceptibility is equal to its static counterpart, the system is ergodic}.
However, the problem of solving the dynamical equations of motion for many-body
systems remains a daunting one at best, consequently the validation of the EH
via Kubo's method has been done only for a few systems. 

 With the development of the recurrence relations method by Lee~\cite
{Lee82},  it became possible to obtain general solutions for the Heisenberg
equations  of motion.  In this way it is possible to verify the validity of the EH~\cite{Lee01,Lee06}.We have avoided this problem using the asymptotic limit of the correlation function, which allows us to verify the EH without a full solution of the dynamical equations of motion. For that, we just need the memory, which is our starting point. In this work we try to narrow the difference between the methods, i.e. we show that both methods yield the same result.

\emph{ Ergodicity -- }  Recently, Lee~\cite{Lee06} proposed that the
validation of the EH is subject to the condition 

\begin{equation}
0< \tau=\int_{0}^{\infty }R(t)dt<\infty ,  \label{EH}
\end{equation}
where $\tau$ is a relaxation time we introduce here for further use, $R(t)$ is the renormalized correlation function for the dynamical
stochastic variable $A(t)$. It is given by 
\begin{equation}
R(t)=\frac{\langle A(t)A(0)\rangle }{\langle A(0)^{2}\rangle }  \label{R}
\end{equation}
and here $\langle A(t)A(0)\rangle $ is the ensemble average.  The finite value of the integral is the Lee condition~\cite{Lee06} for the EH be valid.
Note that a necessary condition for the validity of Eq. (\ref{EH}) is that the mixing condition (MC) holds, i.e. 
\begin{equation}
\lim_{t\rightarrow \infty }R(t)=0,  \label{MC}
\end{equation}
otherwise the integral would diverge.  Consequently the MC is necessary for the EH, see Costa \etal~\cite{Costa03}.
The condition imposed by Lee is sufficient for the validity of the EH. However, there are situations where it may be too restrictive and this condition can be weakened. We shall prove here that in anomalous diffusion the MC is a sufficient condition.

Anomalous diffusion can be well described  by a generalized Langevin equation (GLE) of
the Mori form~\cite {Costa03,Morgado02}. From the GLE
it is possible to obtain 
\begin{equation}
\frac{dR(t)}{dt}=-\int_{0}^{t}\Pi (t-t^{\prime })R(t^{\prime })dt^{\prime },
\label{GR}
\end{equation}
where $\Pi (t)$ is the memory. The Laplace transform of this equation yields 
\begin{equation}
\widetilde{R}(z)=\frac{1}{z+\widetilde{\Pi }(z)},  
\label{Rz}
\end{equation}
where the tilde indicates Laplace transforms.
Consider 
\begin{equation}
\lim_{t\rightarrow \infty }R(t)=\lim_{z\rightarrow 0}z\widetilde{R}
(z)=\lim_{z\rightarrow 0}\frac{z}{z+\widetilde{\Pi }(z)},  \label{Rlim}
\end{equation}
where we have used the final value theorem~\cite{Spiegel65}. Note that the inverse Laplace transform can be obtained analytically only in a few cases. However, since we know $\tilde{\Pi}(z)$
the limit can be obtained  even without an explicit solution for $R(t)$.
For 
\begin{equation}
\widetilde{\Pi }(z\rightarrow 0)\sim z^{\nu },  \label{Gaz}
\end{equation}

the MC is not valid for $\nu\geq1$. For diffusive processes with the mean-square displacement given by $\langle x^{2}(t)\rangle \propto t^{\alpha}$, the exponent is~\cite{Morgado02,Costa03} $\alpha =\nu +1$. For $\nu=-1$ we have $\alpha=0$, i.e. no diffusive process. This case needs special analysis: let $\nu=-1$, i.e. $\tilde{\Gamma}(z)= K/z$. The inverse Laplace transform gives $\Gamma(t)=K$. A constant value will produce a harmonic oscillator term which localizes the particle, i.e. no diffusion, in agreement with $\alpha =\nu +1$. From Eq.~(\ref{Rz}), we get that $R(t)=\cos(\sqrt{K} t)$. We can see that this correlation function has an infinite relaxation time, and violates the MC and EH. Consequently, mixing and ergodicity in diffusive processes will hold only
for $-1<\nu<1$, which correspond to $0<\alpha<2$. For $\nu= 0$ (normal diffusion), we can define a finite relaxation time by Eq.~(\ref{EH}). Strictly speaking, one can define a relaxation time only for normal diffusion. In that case, with broad-band noise, the correlation function relaxes as $\exp(-t/\tau)$~\cite{Vainstein06a}. We shall analyze here the extreme cases where the MC condition is violated.

{\em Slow Relaxation Phenomena -- } We shall restrict ourselves to
diffusive phenomena in order to compare our results with those of Lee. The
majority of systems that violate ergodicity present some form of slow
dynamics. For example Mukamel \etal~\cite{Mukamel05} show that Ising systems
with long-range interactions exhibit a relaxation time which diverges
logarithmically with system size.

Another example of violation of the MC, which again implies the violation of the EH, is the ballistic motion, which we shall discuss in the next
section. Let us now address the superdiffusive motion i.e. $0<\nu <1$.

\begin{figure}[htbp]
\centering
\includegraphics[height=8cm,width=6cm,angle=270]{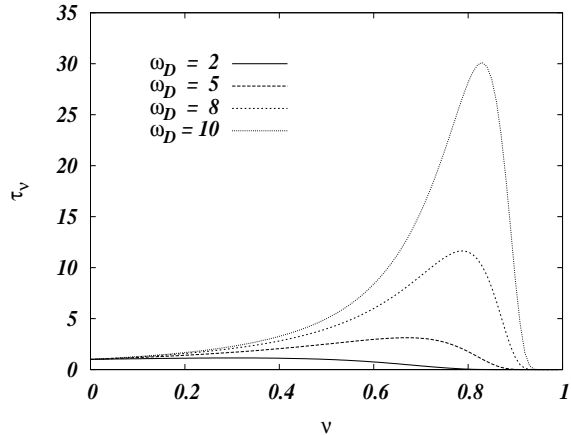}
\caption{Plot of the relaxation time $\tau_{\nu}$ as a function of the index $\nu$ for values
of $\omega_D/\gamma$ equal to $2$, $5$, $8$, and $10$. Note the growth of the maximum as the ratio $\omega_D/\gamma$ increases.}
\label{fig:tau_nu}
\end{figure}

As we have mentioned before, relaxation times exist only for normal diffusion, However, for superdiffusion, Eq. (\ref{GR}) in the limit given by Eq. (\ref{Gaz}), yields a Mittag-Leffler function of the form 
$E_{1-\nu }(-(t/\tau _{\nu })^{1-\nu}),$ 
which displays a transient behavior from a stretched exponential $ \exp \left[ -\left( t/\tau_{\nu }\right) ^{1-\nu }\right] $
to a power law $\left( t/\tau _{\nu-1 }\right)
^{\nu -1}$. In order to have an idea of the magnitude of $\tau$, we need to know the behaviour of $\widetilde{\Pi}(z)$ for small $z$. This depends on the nature of the noise in the stochastic process~\cite{Vainstein06}
For a noise density of the form 
\begin{equation}
\rho (\omega )=\left\{ 
\begin{array}{r}
\frac{2\gamma }{\pi }\left( \frac{\omega }{\omega _{D}}\right) ^{\nu },
\text{\qquad if }\omega <\omega _{D} \\ 
0,\text{\qquad otherwise,}
\end{array}
\right.   \label{noise}
\end{equation}
with $\omega _{D}$ as a Debye cutoff frequency, it is possible to compute the memory~\cite{Morgado02,Costa03} using the Laplace transform,
and its low $z$ limit, to obtain the coefficient of Eq.~(\ref{Gaz}). The transient time is $\tau _{\nu }= \tau_0\left[\left(\gamma/\omega_D\right)^\nu \sec \left((\nu\pi)/2\right)\right]^{1/(\nu -1)}$.
   In Fig.~\ref{fig:tau_nu} we plot the transient time $\tau_{\nu}$ as a function of $\nu$ for several values of $\omega_D/\gamma$. For normal diffusion, $t_0=1/\gamma$, it is equivalent to the relaxation time. Notice that the maximum increases with $\omega_D/\gamma$. For broadband noise $\omega_D/\gamma\gg 1$, the transient time becomes very large as $\nu$ approaches $1$.

{\em Ballistic Motion -- }   In
nature, normal diffusion and subdiffusion are prevalent, as can be
observed in most conductors~\cite{Dyre00}. However, very recently in the
history of conductivity investigations, superdiffusive and even ballistic
motion have been produced in laboratories. 
This introduces a new and important field of investigation~\cite{Frank98,Poncharal02,Bellani99,Bellani00,Oliveira05,Vainstein05}. 
 Indeed, we can find reliable
reports on ballistic conductivity in carbon nanotubes~\cite
{Frank98,Poncharal02}, in semiconductors~\cite{Hu95}, and in semiconductor
superlattices with intentionally-correlated disorder~\cite
{Bellani99,Bellani00}.  We discuss here the violation of MH and EH in ballistic motion.

For ballistic conduction  $\tilde{\Pi}(z) \propto cz$, where $c$ is a number without dimensions. The limit becomes
\begin{equation}
\lim_{t\rightarrow \infty }R(t)=(1+c)^{-1} \ne 0,
\end{equation}
and the MC is violated. This implies that the EH and the FDT are  violated~\cite {Costa03}.

  Briefly, for some processes like anomalous diffusion the Lee condition  Eq.~(\ref{EH}) for the EH is too restrictive and in order to have the EH, we need only the MC. Moreover, the integral is finite only for normal diffusion, $\alpha=1$. For anomalous diffusion in the range $0<\alpha<2$, the integral is either null ($0<\alpha<1$) or  infinite ($1<\alpha<2$).
In all these cases, the MC, the EH and the FDT are valid. At the extreme limit $\nu=-1$, or $\alpha=0$, the system behaves as a localized harmonic oscillator and violates the MC, the EH and the FDT; for $\alpha=2$, ballistic
motion, the system also violates MC, and consequently the EH, and the FDT fail.

\emph{Conclusion -- } In this work we revisited the problem of the
validity of the EH and that of the MC, and we obtained an agreement between our previous result~\cite
{Costa03} and recent results using the recurrence relations method~\cite
{Lee06}. However, for anomalous diffusion our condition can be less restrictive than that of Lee. Since diffusion is a main phenomenon in physics we use that as our
starting point. The method used by Lee~\cite{Lee82,Lee01,Lee06} is quite general and applies to any response function. In this context, the
equivalence between this and our results strengthens both. The violation of
the EH is exhibited for ballistic and harmonic motions. We have not focused deeper on real
complex systems; we chose to follow easy-to-understand concepts where limits
can be analytically obtained. This gave us a good framework for analyzing
more complex structures. Nonlinear dynamics is a field which deserves much
attention; in particular, the coalescence of trajectories has been
intensively studied in the last few years~\cite{Longa96,Ciesla01,Boccaletti02}. The restriction of the degrees of freedom
there may confirm the hierarchy exposed here. We also expect that new mathematical methods~\cite{Dorea06} may bring alternative proofs to the problem.

\emph{Acknowledgments --}  We thank professor Howard Lee for
letting us read his manuscript prior to publication. This work was supported by
the Brazilian agencies: CAPES, CNPq, FINEP, and FINATEC.

\bibliographystyle{unsrt}

\end{document}